\documentclass[]{aa}
\usepackage[T1]{fontenc}
\usepackage[utf8]{inputenc}

\usepackage{graphicx}
\usepackage{natbib}
\usepackage{ulem}
\usepackage{textcomp}
\usepackage{gensymb}
\usepackage{longtable}
\usepackage{threeparttable}
\usepackage{multicol}
\usepackage{multirow}
\usepackage{float}
\usepackage{todonotes}
\usepackage[Symbol]{upgreek}
\usepackage{amsmath}
\usepackage{etoolbox}
\usepackage{txfonts}
\usepackage{url}
\usepackage{xcolor}

\usepackage[most]{tcolorbox}
\usepackage{hyperref}
\usepackage{xcolor}
\newcommand{\enzo}{{\it {\small ENZO\,}}}

\begin{document}

\title{New constraints on the magnetic field in cosmic web filaments}

\author{N. Locatelli\inst{1,2,3}, F. Vazza\inst{1,2,4}, A. Bonafede\inst{1,2,4},  S. Banfi\inst{1,2}, G. Bernardi\inst{2,5,6}, C. Gheller\inst{2}, A. Botteon\inst{7}, T. Shimwell\inst{7,8}}

\offprints{%
 E-mail: nlocat@mpe.mpg.de}
\institute{Dipartimento di Fisica e Astronomia, Universit\'{a} di Bologna, Via Gobetti 93/2, 40122, Bologna, Italy
\and Istituto di Radioastronomia, INAF, Via Gobetti 101, 40122, Bologna, Italy
\and Max-Planck-Institut f\"ur Extraterrestrische Physik (MPE), Giessenbachstrasse 1, 85748 Garching bei M\"unchen, Germany
\and  Hamburger Sternwarte, University of Hamburg, Gojenbergsweg 112, 21029 Hamburg, Germany
\and Department of Physics and Electronics, Rhodes University, PO Box 94, Grahamstown, 6140, South Africa
\and South African Radio Astronomy Observatory, Black River Park, 2 Fir Street, Observatory, Cape Town, 7925, South Africa
\and Leiden Observatory, Leiden University, PO Box 9513, 2300 RA Leiden, The Netherlands
\and ASTRON, The Netherlands Institute for Radio Astronomy, Postbus2, 7990 AA Dwingeloo, The Netherlands
}

\authorrunning{N. Locatelli et al.}
\titlerunning{Constraints on the cosmic web magnetic field}

\date{Accepted ???. Received ???; in original form ???}

\abstract{
Strong accretion shocks are expected to illuminate the warm-hot inter-galactic medium encompassed by the filaments of the cosmic web, through synchrotron radio emission.
Given their high sensitivity, low-frequency large radio facilities may already be able to detect signatures of this extended radio emission from the region in between two close and massive galaxy clusters. 
In this work we exploit the non-detection of such diffuse emission by deep observations of two pairs of relatively close ($\simeq 10$~Mpc) and massive ($M_{500}\geq 10^{14}M_\odot$) galaxy clusters using the LOw-Frequency ARray (LOFAR). 
By combining the results from the two putative inter-cluster filaments, we derive new independent constraints on the median strength of inter-galactic magnetic fields: $B_{\rm 10Mpc}< 2.5\times 10^2\,\rm nG\,(95\%\, \rm CL)$. 
Based on cosmological simulations and assuming a primordial origin of the B-fields, these estimates can be used to limit the amplitude of primordial seed magnetic fields: $B_0\leq10\,\rm nG$. We advise the observation of similar cluster pairs as a powerful tool to set tight constraints on the amplitude of extragalactic magnetic fields. 
}
\maketitle

\label{firstpage}

\section{Introduction}

At the largest scales of the Universe ($\geq 10$~Mpc), galaxy groups and clusters are connected by elongated distributions of galaxies called filaments  and sheets which are believed to be also permeated by diffuse gas, and possibly by magnetic fields.
Until now, a straightforward and direct detection of inter-galactic medium (IGM) and magnetic field (IGMF) has been prevented by the very low density of the plasma ($n_{\rm IGM}\leq 10^{-4}\,\rm cm^{-3}$) and its  relatively low temperature ($T_{\rm IGM} \leq 10^7$ K). 
However, increasing evidence \citep{2018Natur.558..406N, 2020Natur.581..391M, 2020A&A...643L...2T} is recently confirming the long-lived expectations for the warm-hot gas phase of the IGM (WHIM, with $T_{\rm WHIM}\sim 10^5-10^7$, $n_{\rm WHIM}\sim 10^{-5}-10^{-4}$) to contain up to half of the baryon content at low redshift \citep{1999ApJ...514....1C, 2001ApJ...552..473D, 2021A&A...647A...2R}.

Accretion shocks are believed to reside along and within the filaments of the cosmic web as well as at the outskirts of galaxy clusters. These shocks are expected to amplify the magnetic fields and to accelerate particles up to relativistic energies \citep[][]{2008Sci...320..909R}. 
Their presence might then enable the detection of the WHIM through its synchrotron emission signature at radio wavelengths, and indeed the direct observation of the tip of the iceberg of this diffuse emission has already been obtained at radio frequencies \citep{2019Sci...364..981G, 2020arXiv200809613B}. In these few cases the plasma conditions are still hotter and denser than what expected for the WHIM and the detected emission lays within the clusters virial radii.
Further works in this direction will be helped in the near future by the promising new and upcoming radio facilities (the next generation Very Large Array ngVLA, the Karoo Array Telescope MeerKAT, the square kilometer array SKA-mid) and especially at very low frequencies (LOFAR, the Murchison Widefield Array MWA, SKA-low) where the emission should be brighter up to further out the clusters virial radii thanks to the expected spectral behaviour as $S_\nu \propto \nu^{-1}$ with respect to frequency $\nu$ \citep{2015A&A...580A.119V}.
A way to overcome sensitivity limitations is to quantify the Faraday effect induced by magneto-ionized plasma along the line of sight to a polarized background radio source and build a tomography of the WHIM by means of a grid of background sources (see \citealt{2014ApJ...790..123A, 2016A&A...591A..13V}). A thorough exploitation of this method currently suffers from the lack of large and dense grids of polarized sources, however it is expected to provide important results thanks to the upcoming radio facilities \citep{2018Galax...6..128L}.
Complementary, recent upper limits on the IGMF intensity and scale have been derived from the cross-correlation of diffuse radio synchrotron emission with the underlying galaxy distribution \citep{2017MNRAS.467.4914V, 2017MNRAS.468.4246B}, by cross-correlating the difference in rotation measures of physically related pairs of extended radio galaxies, compared with the one derived from randomly paired and close lobes \citep{2019ApJ...878...92V, 2020MNRAS.495.2607O, 2020A&A...638A..48S} and by stacking full sky low-frequency radio images of close red luminous galaxy pairs \citep{2021arXiv210109331V}.

Why is it important to assess the IGMF properties in the cosmic web at late times?
The magnetic fields in galaxies and galaxy clusters, commonly observed today, arise from strong amplification from efficient MHD small-scale mechanisms \citep{2008Sci...320..909R} which are responsible of a fast saturation of the fields, thus erasing information on their initial conditions, and in turn of their origin \citep{2016ApJ...817..127B}. 
Instead, in the WHIM environment, the amplification of primordial magnetic fields is found in simulations to be mainly driven by the field compression as its lines freeze into the plasma plus the contribution of small scale shocks. These mechanisms do not bring the field to saturation and provide a tool of assessing the history and original conditions of the field by means of the level of magnetisation observed today \citep{2014MNRAS.445.3706V, 2015A&A...580A.119V,2018SSRv..214..122D}. 
For the above reasons it is crucial to constrain the magnetic field in the WHIM in order to determine the original scenario for the large scale magnetic field origin and evolution in the Universe. 

Cosmological MHD simulations predict the intensity of the IGMF at low redshift to range within 1 and 100 nG \citep{1999A&A...348..351D, 2005ApJ...631L..21B, 2017CQGra..34w4001V}. 
In this paper we introduce a novel method for a robust inference of an upper limit on the initial $B_0$ and current $B$ values of the IGMF within the large-scale filaments of the cosmic web. 
The method explores the amount of diffuse emission detected at $144$~MHz with the LOw Frequency ARray (LOFAR) telescope along the direction connecting two different pairs of galaxy clusters.
\begin{figure}
    \centering
    \includegraphics[width=\linewidth]{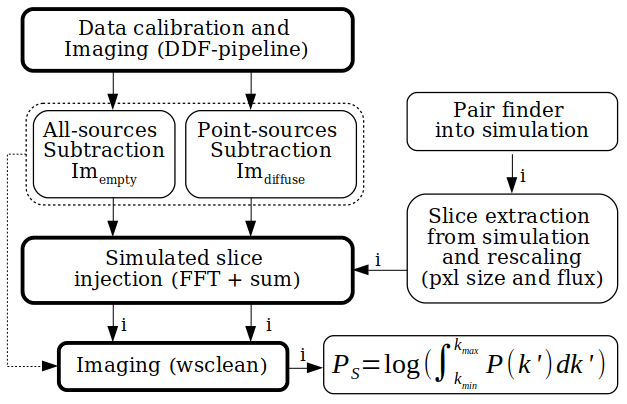}
    \caption{Diagram of the method outlined in Sect.~\ref{sec:method}. Thick boxes highlight the most computationally expensive steps. Links labelled with the letter "i" are computed iteratively over the simulated pairs.}
    \label{fig:scheme}
\end{figure}
We outline the method used to explore the upper limits on the IGMF into cosmological filaments in the following Sect.~\ref{sec:method}; we show its results in Sect.~\ref{sec:results} and discuss their assumptions and implications in Sect.~\ref{sec:discussion}; we draw our conclusions in Sect.~\ref{sec:conclusion}. 
Throughout this work we assumed a $\Lambda$CDM cosmological model, with baryonic and dark matter and dark energy density parameters $\Omega_{\rm BM}= 0.0455,\,\Omega_{\rm DM}=0.2265,\, \Omega_\Lambda=0.728$ respectively and a Hubble constant $H_0= 70.2 \rm km \,s^{-1}\, Mpc^{-1}$. 
\begin{figure*}
    \centering
    \includegraphics[width=0.49\linewidth]{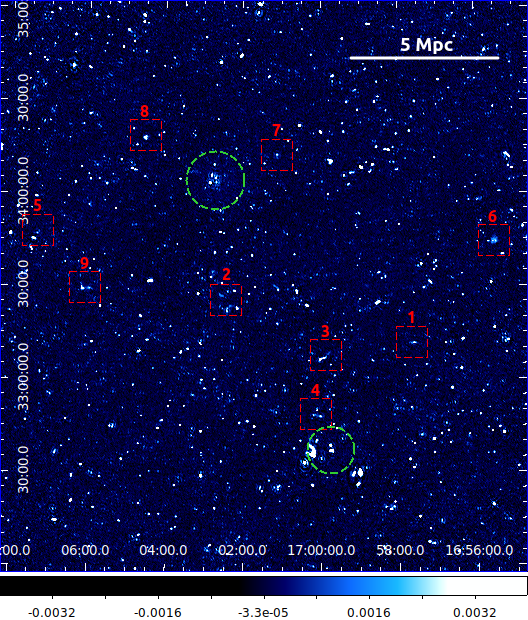}
    \includegraphics[width=0.49\linewidth]{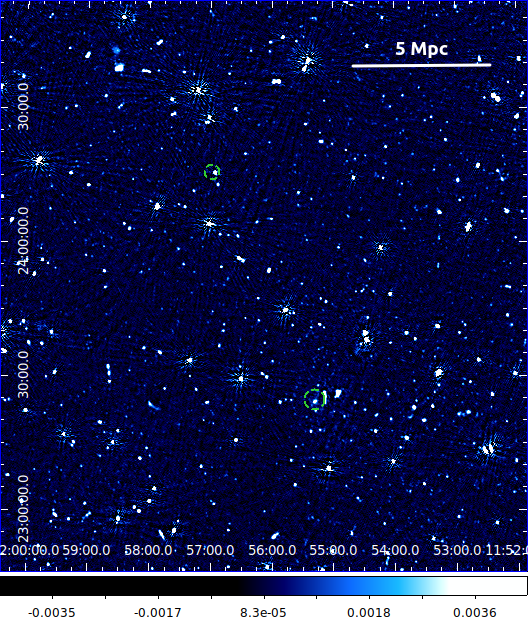}
    \caption{LOFAR low-resolution ($20\arcsec$) images at 150 MHz of the cluster pairs RXC\_J1659-J1702 (left panel) and RXC\_J1155-J1156 (right panel). The dashed circles are centered on the clusters and have corresponding radius $R_{500}$. The red dashed boxes in the left panel have been zoomed in Fig.~\ref{fig:diffuse_patches}. We indicated the 5 Mpc unit at the redshift of the pairs $z=0.10, \,0.14$ (left and right panels). Color-bars are in units of Jy~beam$^{-1}$.}
    \label{fig:pairs_print}
\end{figure*}

\section{Method} \label{sec:method}

In order to look for large scale emission from the cosmic web, we observed pairs of galaxy clusters and the putative inter-cluster filaments connecting them. The cluster pairs have been selected from the Meta-Catalogue of X-Ray Detected Clusters of Galaxies \footnote{ http://heasarc.gsfc.nasa.gov/W3Browse/all/mcxc.html} \citep[MCXC][]{2011A&A...534A.109P} by applying cuts in declination ($\delta\geq 10\deg$), redshift ($z\leq 0.3$) and maximum angular separation ($\theta \leq 5\deg$). These values have been tailored to the proposed LOFAR observations. 
The two most promising pairs that, according to cosmological simulations, maximise the probability of a physical connection between the clusters in terms of total mass and separation (real and projected), have been proposed and observed at the LOFAR during Cycle 9 (Proposal Id:LC9\_020).
The most important properties of the two observed pairs of clusters are given in Table~\ref{tab:cluster_properties}.

In a nutshell, after calibrating, imaging and removing contaminating sources from the LOFAR data, we  quantify the confidence of having observed (or not) diffuse emission from the inter-cluster filaments by injecting simulated diffuse emission produced by large scale ($\geq$Mpc) accretion shocks, into the original radio visibility data (uvw) for a large ($\mathcal{O}(100)$) subset of simulated filaments/cluster pairs and by imaging them as done for the real observations.

In this section we provide further details on the analysis performed on the actual observations, the simulated data set prepared for the injection and the injection procedure (also sketched by the diagram in Fig.~\ref{fig:scheme}).

\subsection{LOFAR radio observations}

We have observed the two cluster pairs RXCJ1659.7+3236-RXCJ1702.7+3403 (hereafter RXC\_J1659-J1702) and RXCJ1155.3+2324-RXCJ1156.9+2415 (hereafter RXC\_J1155-J1156) using LOFAR. The fields containing these two targets have been co-observed together with two pointings of the LOFAR Two-metre Sky Survey \citep[LoTSS;][]{2017A&A...598A.104S} taking advantage of the multi-beam capabilities of LOFAR. The observing setup of our observations thus follows that of LoTSS, namely 8~hr on-source time book-ended by two 10~min scans on the flux density calibrator in the frequency range 120-168 MHz using LOFAR in \texttt{HBA\_DUAL\_INNER} mode \citep[see][for details]{2017A&A...598A.104S}.
A first calibration and imaging run has been performed adopting the pipelines developed to analyse LoTSS pointings \citep{2017A&A...598A.104S, 2019A&A...622A...1S}, that aim to correct both for direction-independent and direction-dependent effects exploiting {\small PREFACTOR} \citep{2016MNRAS.460.2385W, 2016ApJS..223....2V, 2019A&A...622A...5D}, {\small KILLMS} \citep{2014arXiv1410.8706T,2014A&A...566A.127T,2015MNRAS.449.2668S}, and {\small DDFACET} \citep{2018A&A...611A..87T} softwares. In particular, we have used the improved version of the direction-dependent data reduction pipeline (v2.2\footnote{https://github.com/mhardcastle/ddf-pipeline}, the same used for the forthcoming second LoTSS data release DR2, Shimwell et al. in preparation) to produce images of the full LOFAR field-of-view at the central frequency of 144~MHz at high ($6 \arcsec$) and low resolution ($20 \arcsec$, shown in Fig.~\ref{fig:pairs_print}) using a Briggs weighting scheme (robust=-0.5). We refer the reader to \cite{2021A&A...648A...1T} for a thorough description of the steps performed by the pipeline.
Using the sky models derived from the pipeline, we have subtracted the sources from the uv-data in two different fashions: either we have subtracted  all sources (by means of their clean components) found in the high and low resolution maps or we have subtracted all sources detected in the high resolution image.
The model components have been determined during the high and low resolution images deconvolution making use of the PYthon Blob Detector and Source Finder ({\small PYBDSF}; \citealt{2015ascl.soft02007M}).
The subtraction of the model components has been performed in the visibility domain, corrupting the model components by the direction-independent antenna gains obtained from the calibration.  
We have then produced dirty images from the subtracted data. The images include the residual contribute to the surface brightness resulting from model approximation plus artefacts associated with imperfect model and solutions ($\rm Im_{empty}$) plus patches of faint extended emission in the case in which only sources from the high resolution model have been subtracted ($\rm Im_{diffuse}$). 
The subtracted (dirty-)images $\rm Im_{empty}$ at low resolution ($20\arcsec$) have a rms noise floor of $\sim 160$ and $240 \,\mu \rm Jy beam^{-1}$ for the two fields, respectively. The noise difference is consistent with the amount of flagged (i.e. discarded) data in the two observations. 

\begin{table*}
\centering
\begin{tabular}{l c c c c c c c c c} 
\hline
Cluster name  &  R.A.      & Dec.     & $z$ &   $L_X$ & $M_{\rm 500}$  &  $R_{\rm 500}$ & $\frac{d _{3D}}{(R_1+R_2)}$ & angular separation & $L_{\rm fila}$ \\
                        &        [h,m,s]           &    [$^\circ$,$'$,$"$]        &  &  [erg/s]  &  [$M_{\odot}$] & [$\rm Mpc$] &  & $[^\circ]$ & [Mpc]\\
\hline\hline 
  RXCJ1155.3+2324    & 11 55 18  & +23 24 27  &  0.142   &  $6.04 \cdot 10^{44}$  &  $5.60	 \cdot 10^{14}$   &    1.19 &  7.0 & 0.93 & 14 \\
  RXCJ1156.9+2415     & 11 56 58 & +24 15 29  & 0.139   &  $1.50 \cdot 10^{44}$  &  $2.40 \cdot 10^{14}$  & 0.90 & 7.0 & 0.93 & 14 \\ 
  \hline
RXCJ1659.7+3236 &       16 59 44 &   +32  36 49    &   0.101  & $1.12 \cdot 10^{44}$ & $2.04 \cdot 10^{14}$ &      0.87 & 13.3 & 1.57 & 25 \\
  RXCJ1702.7+3403 &     17 02 42 & +34  03 43   &  0.095    & $4.04 \cdot 10^{44}$  &   $4.49 \cdot 10^{14}$ &  1.01 & 13.3 & 1.57 & 25 \\
\end{tabular}
 \caption{Main parameters of the two pairs of galaxy clusters observed in this work, based on the Meta-Catalogue of X-Ray Detected Clusters of Galaxies \citep[MCXC][]{2011A&A...534A.109P}. In the last three columns, we give the 3-dimensional distance of between the two cluster centres, normalized by the radius of the two clusters), the angular separation of the two cluster centres, and the 3-dimensional length of the filament (considering the cluster to cluster distance).}
 \label{tab:cluster_properties}
\end{table*}

\subsection{Cosmological simulations}

We have extracted the simulated inter-cluster filaments from the suite of simulations of the cosmic web properties described in \cite{2019A&A...627A...5V} performed with the cosmological MHD code \enzo \footnote{www.enzo-project.org} \citep{2014ApJS..211...19B}. They consist in a comoving $100^3\,\rm Mpc^3$ box with a uniform grid of $2400^3$ cells (and $2400^3$ dark matter particles) with linear (comoving) resolution of 41.6 kpc per cell and dark matter mass $m_{dm}=8.62\times 10^6\,M_\odot$ per dark matter particle. 
Magnetic fields have been initialized at $z=45$ as a uniform background of of $B_0=0.1$~nG and evolved at run-time using the MHD method of Dedner \citep{2002JCoPh.175..645D}. We note that a uniform initial magnetic field here corresponds to a scale-invariant spectrum in the models used for Cosmic Microwave Background (CMB) analysis \citep{2019A&A...632A..47A,2019JCAP...11..028P}. We also note that the run is non-radiative and does not include any treatment for star formation or feedback from active galactic nuclei (AGN). While to a first approximation, these processes are not very relevant for the radio and X-ray properties of the peripheral regions of galaxy clusters and filaments,\citep[e.g.][]{2019MNRAS.486..981G}, the effect of outflows from AGN and galaxies can mix metals and magnetic fields at least out to the virial radius of clusters \citep[e.g.][]{2018MNRAS.476.2689B}.  Simulations by our group have showed that such effects is negligible on the radio emission on the scale of filaments \citep{2017CQGra..34w4001V}, however only future and more refined simulations including galaxy formation-related effects in filaments, at a much higher resolution, will be able to assess with more certainty whether the limits obtained from observations (such as the one obtained in this very work) can be straightforwardly related to a limit on primordial seed fields.
\begin{figure}
    \includegraphics[width=\linewidth]{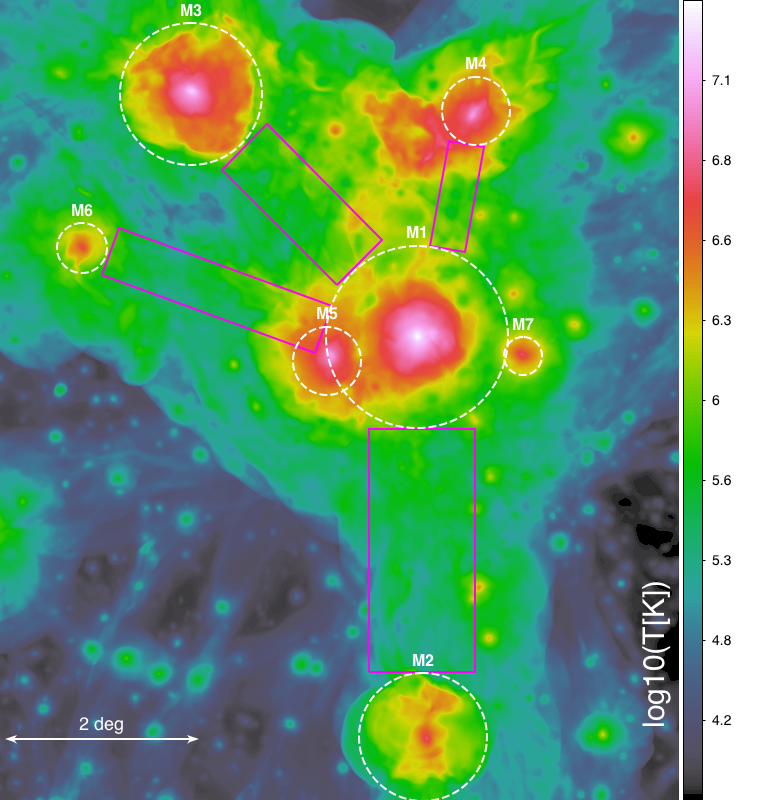}
    \includegraphics[width=\linewidth]{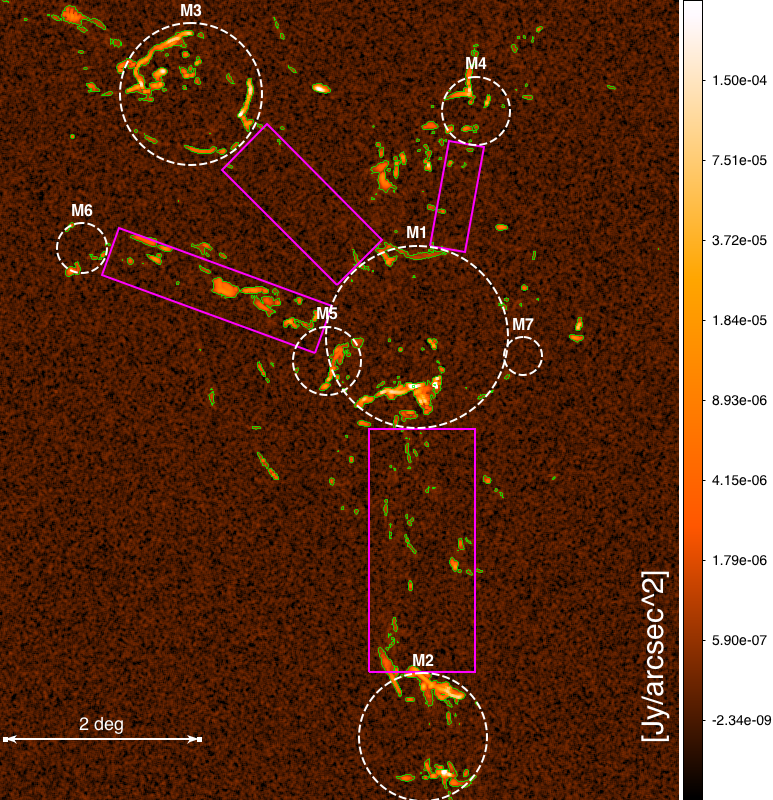}
    \caption{\small Example of mean gas temperature (top) and mock LOFAR-HBA observation (bottom, $\nu=140$ MHz, $25\arcsec$ resolution, 250$\mu\rm Jy\, beam^{-1}$ noised added) from the \enzo  simulation. Circles mark  the virial region of clusters, magenta rectangles mark filaments. The detectable emission ($\geq 3 \sigma$) is marked with green contours. The cluster field is placed at $z=0.1$.}  
    \label{fig:fila_maps}
\end{figure}

\subsection{Synchrotron emission model for cosmic shocks}

We have produced synthetic maps of synchrotron radio emission assuming that diffusive shock acceleration (DSA, e.g. \citealt{2012ApJ...756...97K} and references therein) accelerates a small fraction of thermal electrons swept by structure formation shocks up to relativistic energies, as in \citet{2019A&A...627A...5V}. We have computed the radio emission from electrons in the downstream cooling region of shocks using the model of \citet{2007MNRAS.375...77H} and based on the shocks identified in post-processing in the simulation. The total acceleration efficiency at shocks, $\xi_e(\mathcal{M})$ (with $\mathcal{M}$ the Mach number) is assumed to be the combination of two variables: the kinetic energy flux dissipated onto the acceleration of cosmic rays, $\psi(\mathcal{M})$, and the fraction going into electron acceleration, $\xi'_e$, giving $\xi_e(\mathcal{M})=\xi'_e \cdot \psi(\mathcal{M})$. 
Following \citet{2007MNRAS.375...77H}, the radio emission in the downstream of each shock is directly linked to the power-law energy distributions $N_\gamma \propto \gamma^{-p}$ of electrons  accelerated by the shock front during a cooling time, through the integrated radio spectrum of $I(\nu) \propto \nu^{-s}$, where $s=(p-1)/2+1/2$, with $p=2(\mathcal{M}^2+1)/(\mathcal{M}^2-1)$. With this approach and for the range of $\mathcal{M} \gg 5$ shocks usually found within and around simulated filaments, as well as for the $\leq \rm \mu G$ magnetic fields in filaments  \citep[e.g.][]{2017CQGra..34w4001V}, the radio emission thus scales as $I(\nu) \propto \xi_e B^2 \nu^{-2}$. The baseline model used in this work assumes $\xi_e=10^{-2}$, which is in line with DSA expectations for the maximal acceleration efficiency of relativistic electrons by strong shocks \citep[e.g.][]{2007MNRAS.375...77H,2012ApJ...756...97K,2019SSRv..215...14B} and is also compatible with the modelling of supernova remnants \citep[e.g.][]{2007Natur.449..576U,2019ApJ...876L...8B}. However, we notice that in typically weak ($\mathcal{M} \leq 4$) shocks leading to radio relics in galaxy clusters \citep[e.g.][]{2019SSRv..215...16V}, the acceleration efficiencies implied by the observed relic radio fluxes can be much larger ($\xi_e \sim 0.1-1$, see e.g. \citealt{2019MNRAS.489.3905S} and \citealt{2020A&A...634A..64B}), thus making our maximal value of $\xi_e=10^{-2}$ a conservative one. We notice however that very recent particle-in-cell (PIC) simulations (albeit in 1D and with some limiting assumptions) have derived a $\sim5\%$ electron acceleration efficiency by strong shocks \citep{2020ApJ...897L..41X}. For the remainder of the paper, the reader must thus bare in mind that our limits on $B_{\rm 10Mpc}$ must be accordingly rescaled if a different value of $\xi_e$ is adopted. 

Fig.~\ref{fig:fila_maps} gives an example of filaments connecting a massive cluster to other groups in its surrounding, in an \enzo cosmological simulation. A small but significant fraction of radio emission from shocks running on filaments connecting some of the pairs (e.g. M1-M2, M1-M4 and M1-M6 in Fig.~\ref{fig:fila_maps}) is above the detection threshold in LOFAR-HBA for our baseline model. 
In all cases, the detectable emission comes from relatively small and localised patches, extended a few $\sim 10'$ at most, with irregular shapes.  Detecting the radio signal from cosmic filaments is indeed made challenging by the fact that the detectable fraction is just the tip of the iceberg of the wider 'radio cosmic web', which makes  a morphological classification of the emission often ambiguous. Indeed advanced Deep Learning techniques have been proposed for the detection of the cosmic web in next radio surveys \citep[][]{2018MNRAS.480.3749G}. 
In the following section we use this model to constrain the amplitude of the $\xi_e B^2$ combination based on our real LOFAR observations. 

\subsection{Generation of a mock catalogue of inter-cluster filaments}
For each observed pair of clusters we have selected simulated pairs holding individual cluster masses $M_{500}>10^{13}M_\odot$ and linear (comoving) and projected angular distance of clusters in the pair within $20\%$ deviation from the values of the observed pair (see the last and second-last columns in Table~\ref{tab:cluster_properties}). 
We have obtained 171 simulated cluster pairs selected for RXC\_J1659-J1702 and 139 pairs for RXC\_J1155-J1156. 
The cluster pairs selected above mirror the separation selection criteria of our observations but include less massive clusters that may not involve a physical connection within one pair. We have thus analysed in addition a sub-sample of high-mass clusters ($M_{500}>3 \times 10^{13} M_\odot$) for which a physical connection and the presence of a inter-cluster filament was verified either manually and through a high temperature cut $T_{\rm WHIM}>10^5$~K of the WHIM within the inter-cluster filament. We refer to this sub-sample as {\it best}.

Taken a simulated cluster pair, a box has been drawn and extracted along the direction connecting the pair. The box has been rescaled to match the angular scale and comoving transverse distance indicated by the pixel size and redshifts of the observed clusters, and the intensity of synchrotron emission has been scaled to match its luminosity distance by conserving the total power (see the lower left panel in Fig.~\ref{fig:injection} for an example).

The flux density has been also multiplied by a constant factor $f_B \equiv [B_0/(0.1\,\rm nG)]^2$.
Under the assumption that the amplification of the magnetic field into the simulated filaments is affected by negligible small scale dynamo \citep{2008Sci...320..909R} and that it is thus mainly driven by the adiabatic compression of the magnetic field lines following from flux freezing\footnote{the magnetic flux through a closed loop C enclosing the surface {\bf S} is simply $\rm \Phi_B=\int_S {\bf B} \cdot d{\bf S}$, valid for ideal plasma conditions} into the plasma condensing during structure formation \citep{2017CQGra..34w4001V}, the magnetic field $B$ in the simulation is scalable with respect to $B_0$ and in turn the synchrotron emissivity is scalable with respect to $f_B\propto B_0^2$ (assuming a constant $\xi_e = 10^{-2}$).

\subsection{Injection of model radio emission into real LOFAR images}
\begin{figure*}
    \centering
    \includegraphics[width=0.33\linewidth]{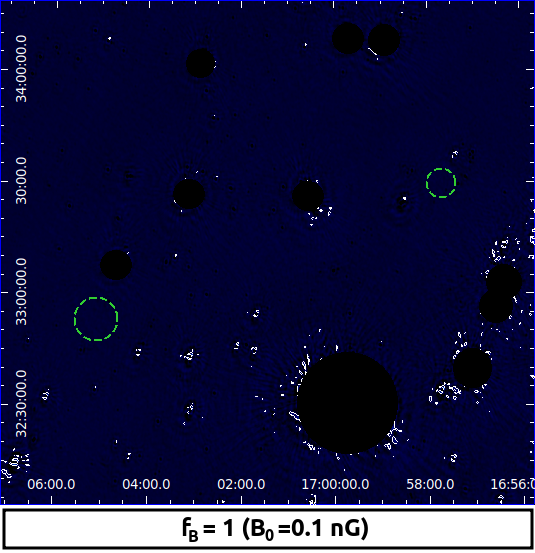}
    \includegraphics[width=0.33\linewidth]{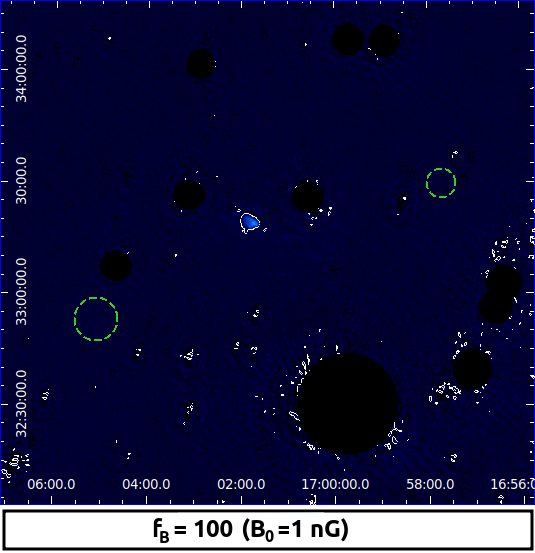}
    \includegraphics[width=0.33\linewidth]{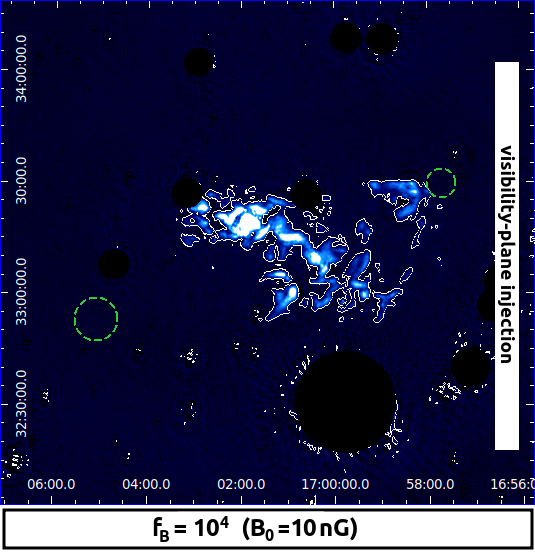}
    \includegraphics[width=0.33\linewidth]{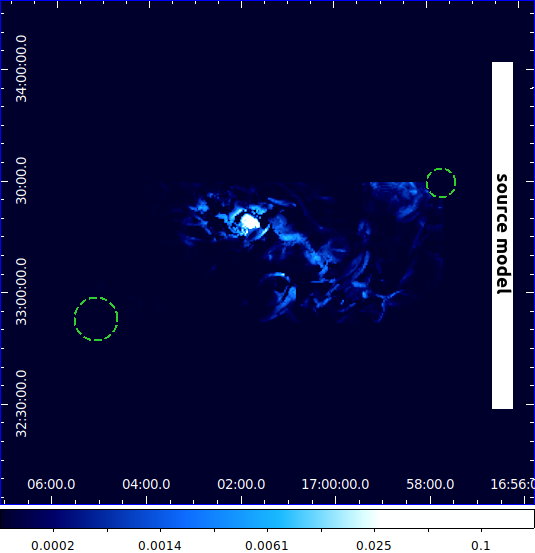}
    \includegraphics[width=0.33\linewidth]{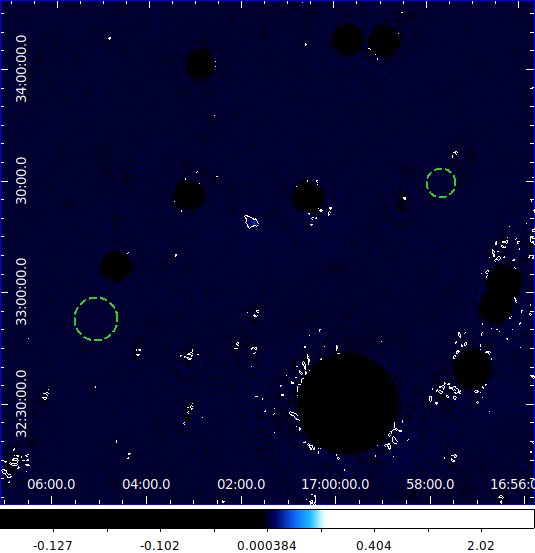}
    \includegraphics[width=0.33\linewidth]{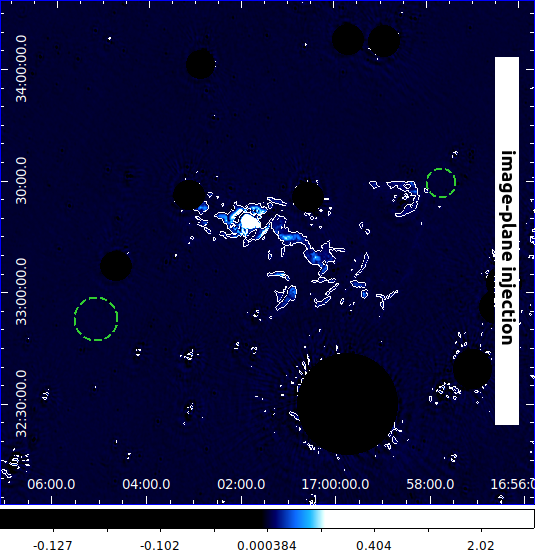}
    \caption{Example of source injection: the model of diffuse emission between a pair of simulated galaxy clusters (masked within their $R_{200}$, indicated by the green dashed circles) found in the simulation with $B_0=0.1$~nG (lower left panel) is multiplied by a factor $f_B=1,\,10^2,\,10^4$ imaged and masked after being injected into the source-subtracted sky visibilities (upper left, upper central and upper right panels respectively), or it is injected through the image-plane (lower central and lower right panels for $f_B=10^2,\,10^4$ respectively). The image with $f_B=1$ in the upper left panel, due to the very low brightness of the model with $B_0=0.1$~nG, results to be equal to the source-subtracted sky image $\rm Im_{empty}$, in which the only features are the residuals from the source subtraction process which fall outside the masks. The white contours have been set to 5 times the rms value in $\rm Im_{empty}$.}
    \label{fig:injection}
\end{figure*}
The rescaled simulated image of each mock pair of galaxy clusters has been injected into the source-subtracted Measurement Set (MS), following the procedure sketched in Fig.\ref{fig:scheme}. In detail, each rescaled image has been first Fourier-transformed, then written into the MS and finally added to the visibilities of the source-subtracted sky using {\small WSCLEAN} \citep{2014MNRAS.444..606O}.
We note that such injection does not take into account direction dependent effects that may act on the MS radio data.
With the same software, the resulting data-set has been imaged and deconvolved with a $20\arcsec$ uv-taper and synthesized beam, and Briggs weighting scheme \citep{1995AAS...18711202B} with robust=-0.25 and 2000 minor cycles (see Fig.~\ref{fig:injection} for output examples). 
For realistic values of the normalisation parameter $f_B$, the detectable emission is fragmented into small and sparse patches, associated with shocks internal to filaments. 
Therefore, we resort to statistical methods to assess the likelihood of each mock image to be compatible with our observed LOFAR fields.
We compute the integral of the image power spectrum $P_S\equiv \log_{10} \left( \int_{k_{\rm min}}^{k_{\rm max}} P(k')\,dk'\right)$ where $P(k')$ is the power spectrum and $k_{\rm min}$ and $k_{\rm max}$ are determined by the image and beam size respectively (see  Fig.~\ref{fig:RXC_J1659-J1702_rob-0.25_1562-1844_Ps_inj+sum} for an example).
The sky model subtraction can leave bright residual artefacts depending on the goodness of the model used. These residuals can be as bright as $\sim 0.1\, \rm Jy\, beam^{-1}$ around point-like sources and they may dominate the integral of the image power spectrum $P_S$. A zero-padding mask has been manually generated for each pair of clusters in order to exclude those artefact from the computation of $P_S$ in all images.

\section{Results} \label{sec:results}

\begin{figure}
    \centering
    \includegraphics[width=\linewidth]{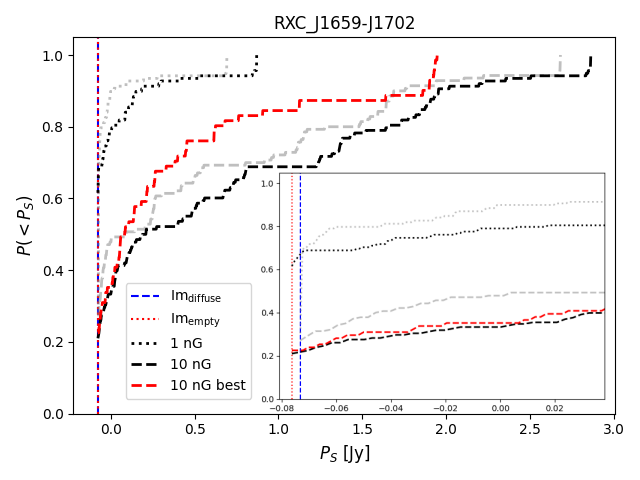}
    \includegraphics[width=\linewidth]{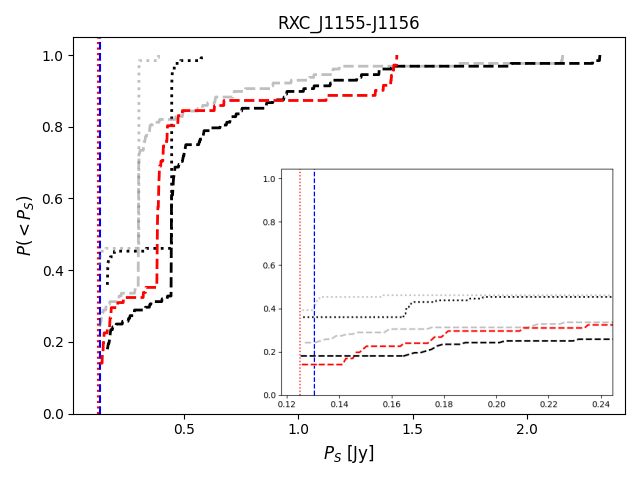}
    \caption{Probability distributions of finding statistics $P_S$ smaller than the values set by the cluster pair RXC\_J1659-J1702 (upper panel) and RXC\_J1155-J1156 (lower panel) for scenarios with $B_0$ as labelled. 
    The vertical lines show the $P_S$ values computed without any injection from $\rm Im_{empty}$ (red dotted), $\rm Im_{diffuse}$ (blue dashed).
    Black and grey lines show results for source injection performed respectively in the visibility and image domains.
    The insets show a zoom on the bins where $P(<P_S)\equiv P(<\tilde{P}_S)$.}
    \label{fig:Prob_PS}
\end{figure}

From the cumulative probability distributions of the statistic $P_S$ resulting from all the source injections, we can access how likely is for a model to provide an expected value smaller than the one recovered from the observations. 
From the image $\rm Im_{empty}$, a $2.5\deg \times 2.5\deg$ square centered on the axis of the cluster pair, in which all the sources (point-like plus extended) have been subtracted, we compute $\tilde{P}_S \equiv P_S(\rm Im_{empty})$. $\tilde{P}_S$ corresponds to the total power in $\rm Im_{empty}$ distributed over all scales from twice the beam size ($k_{\rm min}$) up to half the image size ($k_{\rm max}$). 
All images resulting from injection thus have $P_S$ equal or larger than the one computed for $\rm Im_{empty}$ (red dotted lines in Fig.~\ref{fig:Prob_PS}). 
The statistic $P_S$ resulting from the image in which diffuse emission has not been subtracted $\rm Im_{diffuse}$ are indicated by the blue dashed vertical lines in Fig.~\ref{fig:Prob_PS}. 
Totals of $6$ and $15\, \rm mJy$ of diffuse emission have been found in $\rm Im_{diffuse}$ in excess of $\rm Im_{empty}$ for RXC\_J1659-J1702 and RXC\_J1155-J1156 respectively.
We outline the probabilities for the different models in Table~\ref{tab:probabilities}. The Table reports also results from injection performed in the image plane (instead that in the $uvw$-plane) found in general to produce different probabilities of non-detection with respect to injection through the $uvw$-plane. We discuss this alternative method in Sect.~\ref{sec:discussion}. 

The overall probability of a magnetic field model is simply the product of the probabilities (of the model to produce lower statistics) of the two cluster pairs, since the experiments have been run independently on each pair. We compute these probabilities in the "all" bottom lines in Table~\ref{tab:probabilities}.
A model is more likely to be discarded, when its probability of having a smaller $P_S$ than in our LOFAR observations is very small (or very high alternatively). 

Our main results can be so summarised:
the primordial scenario with a seed magnetic field of $B_0\simeq 10$~nG has a small probability $P(<P_S)\simeq 0.05$ of explaining the small power excess in our observation of the  RXC\_J1659-J1702 and RXC\_J1155-J1156 pairs, we then reject it with a confidence level (CL) of $>95\%$.

The models with a lower seed magnetic fields $B_0<10$~nG yield non-negligible ($\geq0.1$) probabilities to produce a statistic equal to (o smaller than) the one observed. 

By tightening the constraint on individual cluster masses and on the presence of a inter-cluster filament connecting the clusters, the model with $B_0=10$~nG can be rejected with even higher confidence CL$>97\%$.

If any of the patches of diffuse emission observed is produced by shocks in the WHIM, then the $B_0=0.1$~nG model is highly disfavoured, as it is basically unable to produce any detectable emission (i.e the probability in Table~\ref{tab:probabilities} of this model to produce less diffuse emission than what found in $\rm Im_{empty}$ are always $\approx 1$; we note that they have not been plotted in Fig.~\ref{fig:Prob_PS}). 
Although we do not reject this scenario, we consider it implausible (see Sect.~\ref{sec:discussion}).
\begin{table*}
    \centering
    \begin{tabular}{r c c c c c c}
         name & $f_B$ & $B_0$ & \multicolumn{2}{c}{$uvw$-plane} & \multicolumn{2}{c}{image-plane}\\
         {} & {} & [nG] & $\rm Im_{empty}$ & $\rm Im_{diffuse}$ & $\rm Im_{empty}$ & $\rm Im_{diffuse}$ \\
         \hline
         RXC\_J1659-J1702 & 1 & 0.1 & 1 & 1 & 1 & 1 \\
         {} & $10^2$ & 1 & 0.62 & 0.67 & 0.68 & 0.68 \\
         {} & $10^4$ & 10 & 0.21 & 0.21 & 0.27 & 0.27 \\
         *best & {} & {} & 0.23 & 0.23 & 0.24 & 0.24 \\
         RXC\_J1155-J1156 & 1 & 0.1 & 1 & 1 & 1 & 1 \\
         {} & $10^2$ & 1 & 0.36 & 0.36 & 0.42 & 0.42 \\
         {} & $10^4$ & 10 & 0.18 & 0.18 & 0.24 & 0.24 \\
         *best & {} & {} & 0.14 & 0.14 & 0.18 & 0.18 \\
         \hline
         all & 1 & 0.1 & 1 & 1 & 1 & 1 \\
         {} & $10^2$ & 1 & 0.22 & 0.24 & 0.29 & 0.29 \\
         {} & $10^4$ & 10 & 0.04 & 0.04 & 0.06 & 0.06 \\
         *best & {} & {} & 0.02 & 0.03 & 0.04 & 0.04 \\
         \end{tabular}
    \caption{Probabilities of obtaining a statistic $P_S$ lower than the one observed in $\rm Im_{empty}$ and $\rm Im_{diffuse}$, computed from source injection performed in the $uvw$-plane and in the image-plane. 
    }
    \label{tab:probabilities}
\end{table*}

\section{Discussion} \label{sec:discussion}
\begin{figure}
    \centering
    \includegraphics[width=\linewidth]{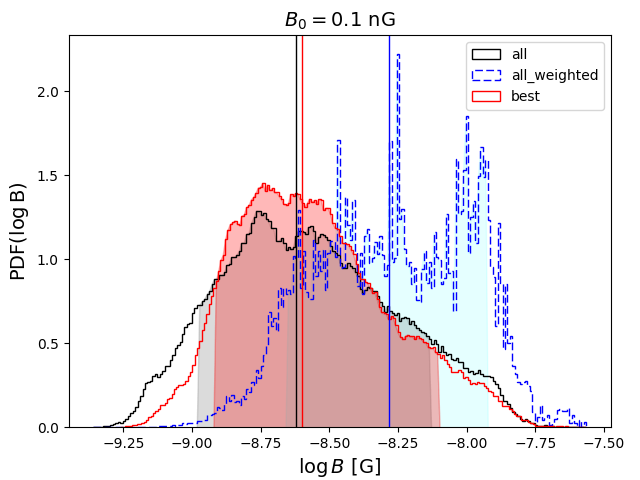}
    \caption{PDFs of the $\log \,B_{\rm 10Mpc}$ field across all the simulated filaments in the $B_0=0.1$~nG model, for all the pairs in the mock sample (black line) and for the {\it best} sub-sample (red line). The dashed blue lines show the $\log \,B_{\rm 10Mpc}$ distribution from all pairs weighted over the pixels emissivity. The filled hatched areas encompass the $10-90$ percentile ranges. The vertical solid lines show the median of the distributions.}
    \label{fig:both_weightedB+B}
\end{figure}
From the original simulation holding $B_0=0.1$~nG, we extract the probability distribution functions (PDF) of the magnetic field values $B_{\rm 10Mpc}$ across the mock filaments selected according to the properties of the observed cluster pairs. We plot the resulting PDF($\log \,B_{\rm 10Mpc}$) in Fig.~\ref{fig:both_weightedB+B}. 
Given the expected lack of dynamo amplification in the WHIM, the magnetic field distributions PDF($\log \,B_{\rm 10Mpc}$) corresponding to the other $B_0$ models can easily be rescaled linearly with the input seed field.
We find a skewed distribution encompassing $B_{\rm 10Mpc}=1.0-7.4$~nG values ($90\%$ confidence range) with median $B_{\rm 10Mpc}=2.5$~nG (equivalent to $\log \,B_{\rm 10Mpc}=-8.6$) for the full sample.
We note that the value of the magnetic field that produces the simulated synchrotron emission lays in the high part of the $B_{\rm 10Mpc}$ distribution, as can be seen from the emission-weighted $B_{\rm 10Mpc}$ distribution in Fig.~\ref{fig:both_weightedB+B}. 
\begin{figure}
    \centering
    \includegraphics[width=\linewidth]{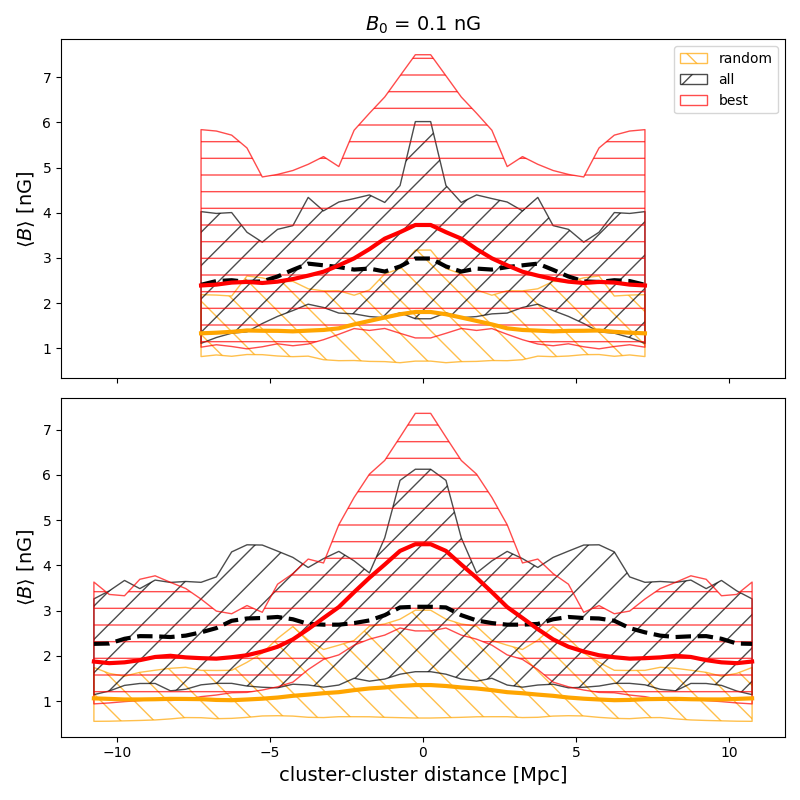}
    \caption{Upper panel: Average profile of mass-weighted magnetic field strength for all inter-cluster filaments (black) extracted to resemble the cluster pair RXC\_J1155-J1156; for the {\it best} sub-sample (red), resulting from the $B_0=0.1$~nG model for a sample of images sampling an equal linear size (15 Mpc) extracted at random locations in the simulation (orange). The solid lines give the median value of the samples while the filled areas encompass the $10-90$th percentiles of the distributions. Lower panel: same plot, for the mock sample extracted for RXC\_J1659-J1702. }
    \label{fig:B_profiles}
\end{figure}
We also give in Fig.\ref{fig:B_profiles} the average profiles of mass-weighted magnetic field strength for all simulated filaments extracted with the procedure above, for the two cluster pairs. On average, the profile of magnetic field is very uniform across $10-20 ~\rm Mpc$, with an  average magnetic field along the line of sight for these objects of $\sim 2-3\rm ~nG$ and a tail of rare and massive filaments that can reach $\sim 10 \rm ~nG$ (Fig.~\ref{fig:B_profiles}). We find that the magnetic field along the direction connecting the clusters is on average enhanced with respect to volume-filling values. The enhancement is significant regardless whether the gas reaches temperatures typical of filaments ($T>10^5$~K) or not, but the enhancement is larger for higher temperatures, as expected from compression of the magnetic field lines during the structures growth \citep[e.g.][]{2016MNRAS.462..448G}.

To interpret the results provided in Fig.~\ref{fig:Prob_PS} and Table~\ref{tab:probabilities}, we postulate three different assumptions that exploit the different type of source-subtraction performed in the analysis, and that can be used to derive different priors from our data (vertical lines in Fig.~\ref{fig:Prob_PS}): 
\begin{itemize}
    \item[I :]{none of the residual diffuse emission after the point-like source subtraction ($\rm Im_{empty}$) is produced by the shocked cosmic web;}
    \item[II :]{all of the residual diffuse emission in excess of $\rm Im_{empty}$ (i.e. $\rm Im_{diffuse}$) is produced by the shocked cosmic web;}
    \item[III :]at least some of the excess diffuse emission present in $\rm Im_{diffuse}$ comes from the cosmic web.
\end{itemize}
We analyse their implications separately in the following paragraphs.

\subsection{Hypothesis I}
Provided that we can fix the $\xi_e$ acceleration efficiency at strong shocks ($\xi_e \approx 10^{-2}$), the assumption that none of the observed emission comes from cosmological shocks (I), produces in principle tighter constraints on $B_{\rm 10Mpc}$ (and $B_0$), since $P(P_S(\rm Im_{empty}))\leq P(P_S(\rm Im_{diffuse}))$ always. 
In practice, the constraints are just slightly tighter due to the small amount of diffuse emission found into $\rm Im_{diffuse}$ with respect to $\rm Im_{empty}$.
Thus, under the first hypothesis that we did not observe the cosmic web emission, by scaling the $B_{\rm 10Mpc}$ distribution obtained from $B_0=0.1$~nG to match the $B_0=10$~nG model (i.e. a factor $\times 100$) we infer an upper limit to the current median IGMF into filaments of $B<0.25\,\rm \mu G$ with $95\%$ confidence (the same confidence level that applies to the rejection of $B_0\geq 10$~nG models of the primordial magnetic field scenario). 
By considering the {\it best} sub-sample of cluster pairs with higher masses and connected by an inter-cluster filament we can further improve the constraints on $B$ by rejecting with a higher CL$>97\%$ the $B_0=10$~nG model.

Not all pairs of clusters are physically connected by a cosmic filaments, and in the absence of a detection in other wavelengths (e.g. via Sunyaev-Zeldovich or X-ray analysis, e.g. \citealt{2019Sci...364..981G}) one needs to resort to linking probabilities as a function of distance, which can be derived from cosmological simulations. For example, early Dark-Matter only simulations estimated that $\sim 80\%$ of pairs in the same mass range of our clusters, and separated by $\sim 15 \rm ~Mpc/h$, are physically connected by filaments  \citep{2005MNRAS.359..272C}. With more recent and resolved simulations, also including gas physics, we can revise this number and tailor it to the exact mass difference and separation of our LOFAR pairs. Using the algorithm outlined in \citet{2021MNRAS.503.4016B}, we reconstructed the network of filaments connecting halos in our simulation, by checking for the actual presence of a matter bridge between pairs of clusters with a $25$ or a $15$~Mpc separation.  This allows us to associate to each LOFAR observation a probability for the presence of a filament,  through the ratio between the number of physical filaments in the simulation ({\it best} sample) over the total pairs of galaxy clusters found at a given distance. This results in 35\% and 65\% probability of having a filament between clusters at $25$ and $15$~Mpc separation, respectively.  
It shall be remarked, that even in the case without an actual gas connection between clusters, the region in between is far from being empty, because massive clusters at a relatively short separation also are indicators of a large cosmic overdensity, which is often associated with the presence of other filaments or threads of the cosmic web along the line of sight, which may account for a non-negligible radio emission. 
This still allows to exclude $B_0 > 10$~nG with high enough confidence. So, the absence of an actual filament does not dramatically affect the validity of the limits inferred from the full sample in this work.

\begin{figure*}
    \centering
    \includegraphics[width=0.32\linewidth]{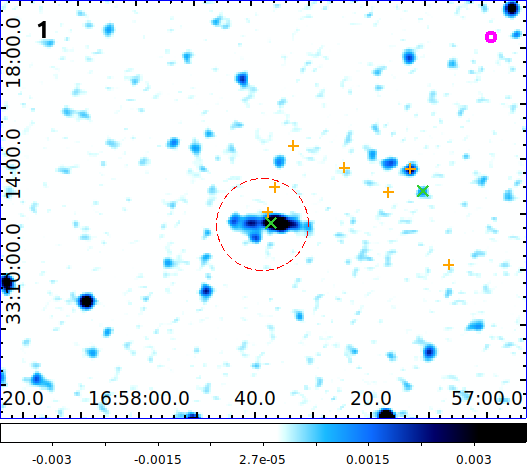}
    \includegraphics[width=0.32\linewidth]{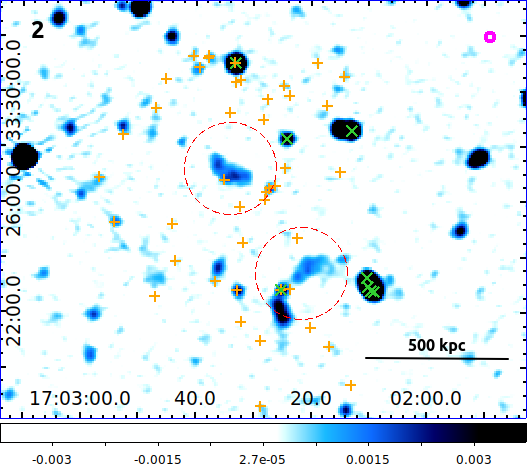}
    \includegraphics[width=0.32\linewidth]{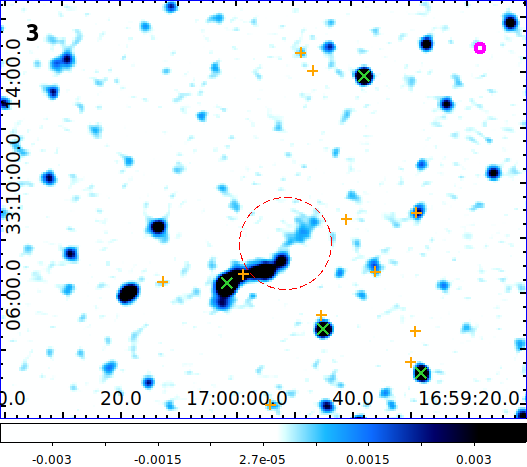}
    \includegraphics[width=0.32\linewidth]{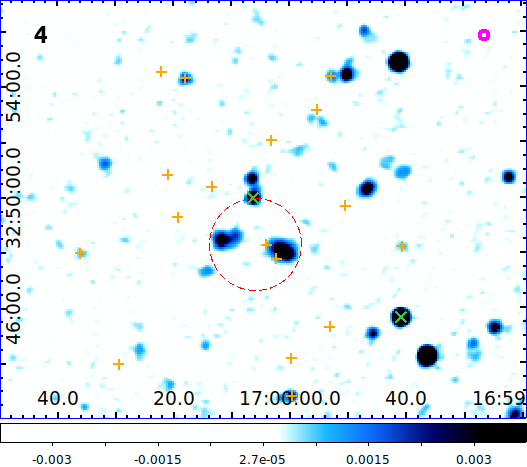}
    \includegraphics[width=0.32\linewidth]{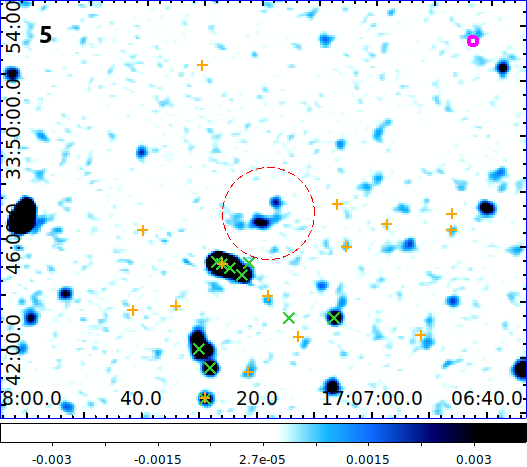}
    \includegraphics[width=0.32\linewidth]{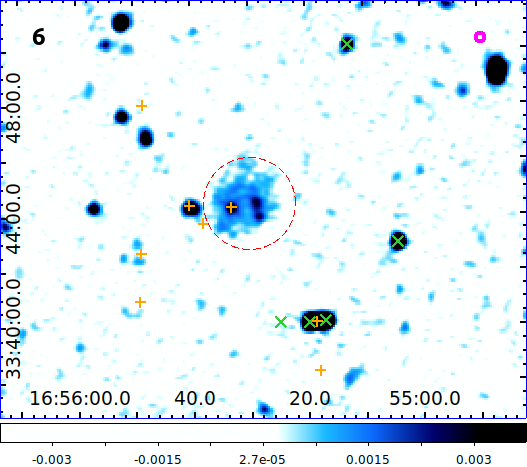}
    \includegraphics[width=0.32\linewidth]{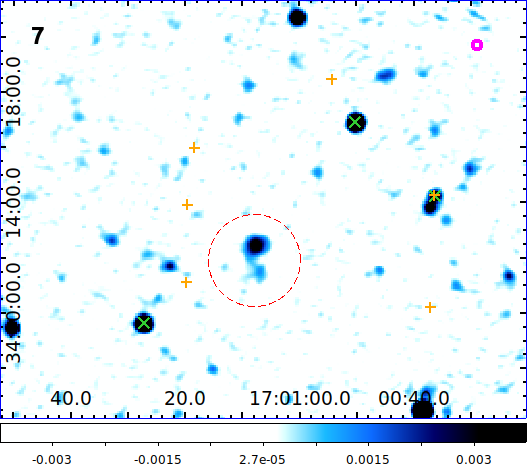}
    \includegraphics[width=0.32\linewidth]{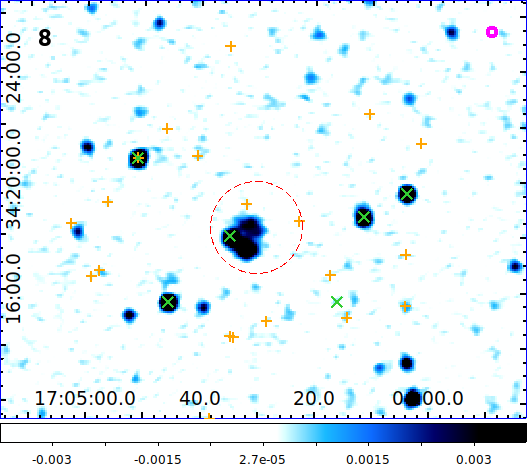}
    \includegraphics[width=0.32\linewidth]{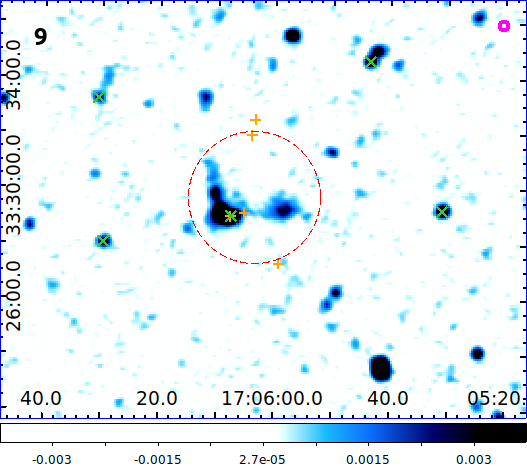}
    \caption{Zoom-in low resolution sky images ($20\arcsec$, see the magenta solid circles) centered over the patches of diffuse emission found in $\rm Im_{diffuse}$ of RXC\_J1659-J1702 as indicated by the dashed red circles. Green X symbols show the position of FIRST sources. Orange crosses show the position of SDSS galaxies with known spectroscopic redshift. Panels are numbered from 1 to 9 going from left to right and top to bottom. $1\arcsec=1.84$~kpc at $z=0.10$, we marked the 500 kpc scale in panel 2 for comparison. Colour bars are in $\rm Jy\,beam^{-1}$.}
    \label{fig:diffuse_patches}
\end{figure*}

\subsection{Hypothesis II}

The assumption that the excess diffuse emission present in $\rm Im_{diffuse}$ with respect to $\rm Im_{empty}$ is entirely due to shocked plasma of the WHIM can be readily tested by looking in detail at the diffuse emission patches which have been detected. 
In Fig.~\ref{fig:diffuse_patches} we present close-up clippings of the diffuse patches found close to the pair RXC\_J1659-J1702, taken from the low resolution LOFAR images (before source-subtraction). They are meant to help in assessing the nature of some of the diffuse emission, indicated by the dashed red circles in the panels of Fig.~\ref{fig:diffuse_patches}.
We also mark with green X symbols the position of sources already known from the VLA Faint Images of the Radio Sky at Twenty-centimeters (FIRST) survey \citep{1995ApJ...450..559B}. 
Most of the diffuse emission is plausibly linked to the lobes of radio-galaxies already detected at higher frequencies.
Panel 7 (panels are numbered from 1 to 9 from left to right, top to bottom) shows what looks like either a radio lobe or an artefact linked to a low-frequency point source.
Panels 2, 5 and 6 instead show diffuse emission which is neither obviously linked to radio galaxies, nor to deconvolution artefacts. 
However, looking at their coordinates, sources 5 and 6 are found to be distant from the axis connecting the clusters, albeit within the imaged portion of the sky around the pair (see also Fig.~\ref{fig:pairs_print}). This makes their physical connection to the putative inter-cluster filament unlikely, even if the shocked cosmic web is expected to fill the space in between clusters in a non-trivial way, as shown in Fig.\ref{fig:fila_maps}. 
Furthermore, the point-like source embedded into the diffuse emission in panel 6 is also found to be at a different redshift with respect to the cluster pair.
For the above reasons, these patches can hardly  be used in the comparison with the simulated inter-cluster filaments. 
The diffuse patches in panel 2 instead embed optical galaxies with redshift $z=0.087,\, 0.093$, consistent with the cluster pair $z=0.095-0.101$, however they are likely dying faint radio lobes, with no FIRST counterpart.
Interestingly, the positions of Sloan Digital Sky Survey (SDSS) sources (orange crosses) in the panels of Fig.~\ref{fig:diffuse_patches} seem to cluster close to the diffuse radio emission. This trend is actually expected if the radio signal is caused by radio galaxies and lobes which show the tendency to be found in clusters or groups of galaxies \citep{1988MNRAS.230..131P, 1993ApJ...404..521A, 1997ApJ...476..489Z, 2011AJ....141...88W}. This behaviour additionally disfavours the link between the excess diffuse emission and the cosmic web.
All in all, since there is no easy way to cross check all the different patches, we still compute the statistics for the most conservative scenario by assuming that the level of observed diffuse emission in excess of $\rm Im_{empty}$ is entirely due to the cosmic web. In this case, the level of confidence associated to the rejection of the same models loosens. However, the models rejected by casting hypothesis II are the same ones resulting from hypothesis I, though with slightly lower or even equal CL. (e.g. for the $B_0=10$~nG {\it best} model the CL for its rejection decreases from $98\%$ to $97\%$ while for the full sample remains unchanged.  
Furthermore, since hypothesis II has been falsified already by the examples described above and shown in Fig.~\ref{fig:diffuse_patches}, hypothesis I is strengthen in favour of hypothesis II and we thus refer to the former in order to draw our conclusions.

\subsection{Hypothesis III}
For completeness, a third additional and interesting way of interpreting our data with respect to the simulation results is the complementary hypothesis to the first one: 
we assume that at least some of the diffuse emission in excess of $\rm Im_{empty}$ comes from the cosmic web. The associated probabilities is then trivially $P(>P_S)=1-P(<P_S)$. In this case we are not interested in the level of diffuse emission in $\rm Im_{diffuse}$, since we want to produce at least the one in $\rm Im_{empty}$.
This scenario, though disfavored, cannot be discarded a priori since this would imply checking (e.g. through cross-correlations) all the different patches of diffuse emission in $\rm Im_{diffuse}$ and prove that all of them are not connected to the emission from the Cosmic Web, it is then instructive to inspect its implications.
Under the assumption that we did see the cosmic web emission at least in part, then the $B_0=0.1$~nG model is ruled out with high confidence $\geq 99\%$ since it is not able to produce any observable emission brighter than the noise level of our LOFAR observation. In this scenario, $B_0>0.1$~nG can be set as a lower limit to the primordial magnetic field intensity and in turn $B>2$~nG as the median value for the magnetic field into filaments today. 

\begin{figure}
    \centering
    \includegraphics[width=\linewidth]{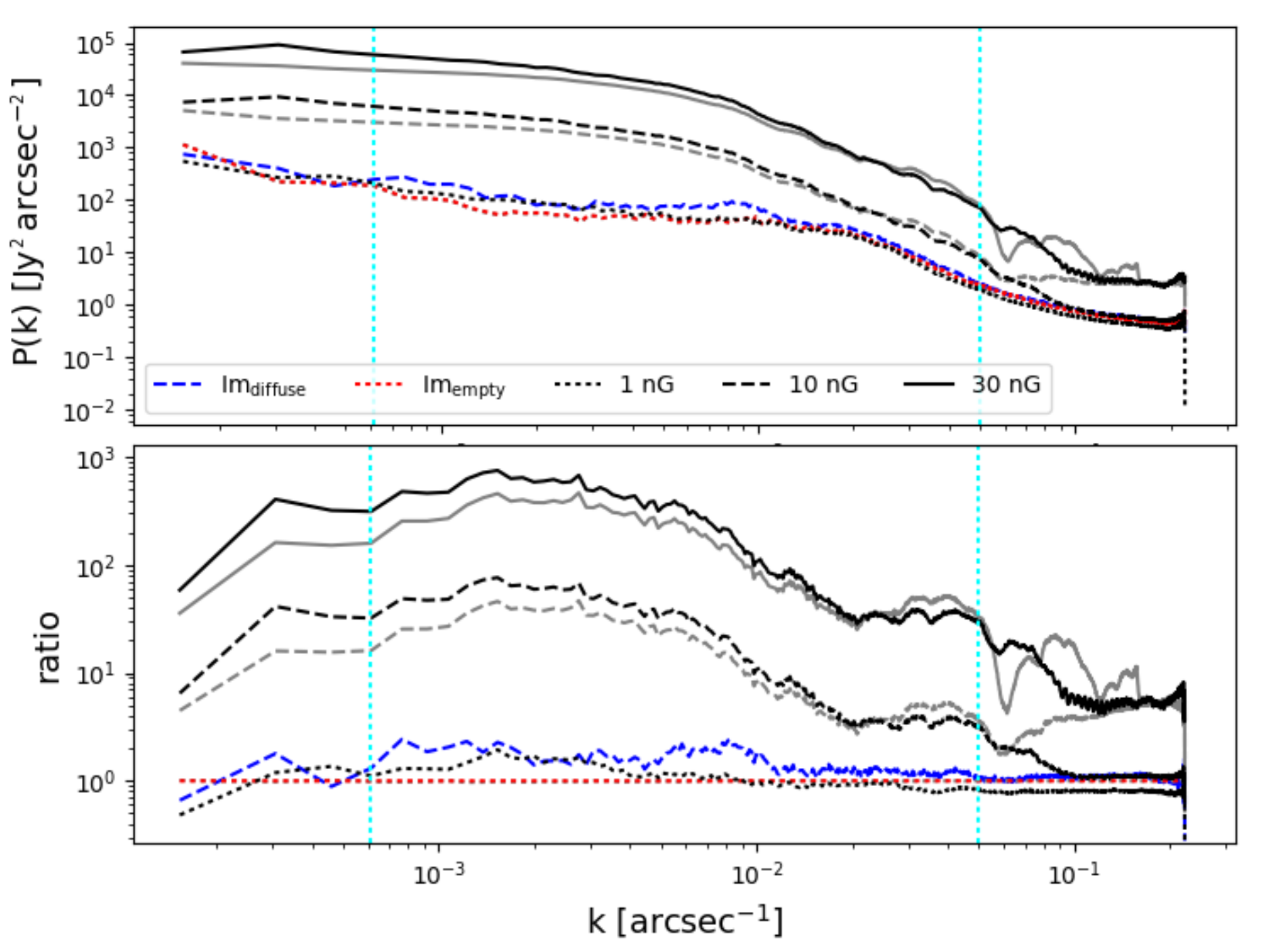}
    \caption{Power spectrum of the injected filament shown in Fig.~\ref{fig:injection} for different $B_0$ models, as labelled. The vertical cyan dotted lines show the integration scale limits used to compute $P_S$. They correspond respectively to about half the largest scale in the image $k^{-1}=2173\arcsec \sim483$~pixels and $k^{-1}=40\arcsec$ (corresponding to twice the synthesized beam FWHM scale).
    Black lines show $P(k)$ resulting from the source injection into $uvw$ visibilities whereas grey lines show the result from the image-plane addition of the simulated image onto $\rm Im_{empty}$. The lower panel shows the same power spectra as in the upper panel, divided by the $\rm Im_{empty}$ line.}
    \label{fig:RXC_J1659-J1702_rob-0.25_1562-1844_Ps_inj+sum}
\end{figure}
\subsection{Visibility vs. image plane injection}
While checking that the source injection procedure (presented in Sect.~\ref{sec:method} and sketched in Fig.~\ref{fig:scheme}) is actually needed in order to derive robust limits on $B_{\rm 10Mpc}$ and $B_0$ we also demonstrate that the method is essential to interpret observations in details by means of the outcome of simulations when dealing with radio data.
With this respect, we have produced the same statistic $P_S$ for the simulations directly added to the residual image $\rm Im_{empty}$ in terms of simple image sum, rather than following the central $\rm FFT + write + sum$ procedure involving visibilities. 
This procedure is much simpler and faster (shortening the computing time of a factor $\sim 600$). 
In Fig.~\ref{fig:RXC_J1659-J1702_rob-0.25_1562-1844_Ps_inj+sum} we plot the power spectra resulting from the injection in the RXC\_J1659-J1702 field of one source as example (images of the same source are shown in Fig.~\ref{fig:injection}) in order to inspect differences between the injection through the $uvw$- (black lines) and image-plane (grey lines), for the different $B_0$ models.
As can be seen by comparing the black and grey lines, when the injection is performed within the image-plane the level of simulated emission at large scales is generally underestimated. As a consequence the models are consistent with the data with different probability up to $\pm 30\% P(<P_S)$ (see the values in  Table~\ref{tab:probabilities}). 
We interpret the difference in the results as due to the lack of model convolution with the instrument's Point Spread Function (PSF). In addition, the lack of convolution of the emission with a visibility weighting scheme able to maximise the evidence for extended diffuse emission into the data may also play a similar role.
As far as an upper cut on the scales of the emission (corresponding to a lower bound on the baseline length in radio interferometers) is taken into account, and detailed power spectrum information does not constitute the largest budget of uncertainty in one analysis (in our case is the scatter in the properties of the -unknown- inter-cluster WHIM), the image sum is a much faster approach than the source injection through the $uvw$-plane, however it shall be used with caution as results are biased by a different sampling of the scales. The strength of the bias depends either on the sampling (window) function and the source power spectrum.

As a final caveat, our analysis assumes that for strong shocks in and around filaments, the acceleration efficiency of electrons is the one suggested by DSA, i.e. $\xi_e \sim 10^{-2}$. This assumes, in turn, that despite the rather low particle density and magnetisation, shocks can form and undergo particle acceleration similar to what is already observed for the outer regions of galaxy clusters in form of giant radio relics \citep[see][for a review]{2019SSRv..215...16V}. Moreover, our analysis assumes that the acceleration of electrons at shocks can proceed independently on the obliquity between the upstream magnetic field and the shock normal. 
However, recent numerical works by \citet{2020MNRAS.496.3648B} have shown that shocks surrounding the cosmic web are more often quasi-perpendicular than random chance, as an effect to the peculiar gas velocity flow following the formation of filaments. In this case, the vast majority of shocks in filaments are quasi-perpendicular and thus likely to be suitable for efficient electron acceleration \citep{2020ApJ...897L..41X}. 
Furthermore, \citet{2017ApJ...843..147M} have recently reported a significant electron acceleration by the strong  quasi-parallel shock while crossing the Saturn bow shock by the Cassini space mission, i.e. in plasma conditions similar to the intra-cluster medium. The acceleration seems to occur in the portion of the shock where upstream cosmic-ray streaming instabilities generate perpendicular small-scale magnetic field components, leading to  particle acceleration. 

\section{Conclusions} \label{sec:conclusion}

In this work, we attempt for the first time to combine dedicated LOFAR-HBA observations of inter-cluster filaments and numerical simulations of the magnetic cosmic web, in order to derive upper limits on the magnetisation of the WHIM. 

While our LOFAR observations do detect patches of diffuse emission of unclear origin their morphology does not allow us to firmly associate the origin of the most prominent ones to the cosmic web. 
However, the presence of a faint diffuse large scale excess in comparison with numerical models allows us to derive inferences on the average magnetisation of such filaments, and possibly on the allowed initial amplitude of primordial seed magnetic fields. 
As a main outcome of our work, by fixing $\xi_e=0.01$ for strong shocks, we derive an upper limit for the median magnetic field strength in filaments connecting massive galaxy clusters:  $B_{\rm 10Mpc}<0.25\,\rm \mu G$.
Based on the dynamical evolution of magnetic fields given by present simulations (which is mostly dominated by simple compression of magnetic field lines), this also implies an upper limit of  $B_0<10$~nG on the amplitude of primordial seed fields.
The estimates above rely on the assumption that our observations constitute non-detections of diffuse emission from the cosmic web (hypothesis I).

As a mutually exclusive interpretation of our data, if some of the detected emission partially came from the shocked WHIM (hypothesis III), this would imply a median magnetic field of order of $B_{\rm 10Mpc}\geq $~nG  (see e.g. Fig.~\ref{fig:both_weightedB+B}).
This would be an important outcome as it would also possibly indicate  primordial magnetic fields with intensity $B_0\geq 0.1$~nG. 

The hypothesis that all of the excess diffuse emission detected in our maps came from the cosmic web (hypothesis II) can be easily rejected even by a visual check of some of the sources.
Given the uncertainties connected to our method and the limited statistics of 'detections' in our sample, we favour the first interpretation of our results (hypothesis I, setting $B_{\rm 10Mpc}<0.25\,\rm \mu G$ and $B_0<10$~nG).

\begin{figure}
    \centering
    \includegraphics[width=\linewidth]{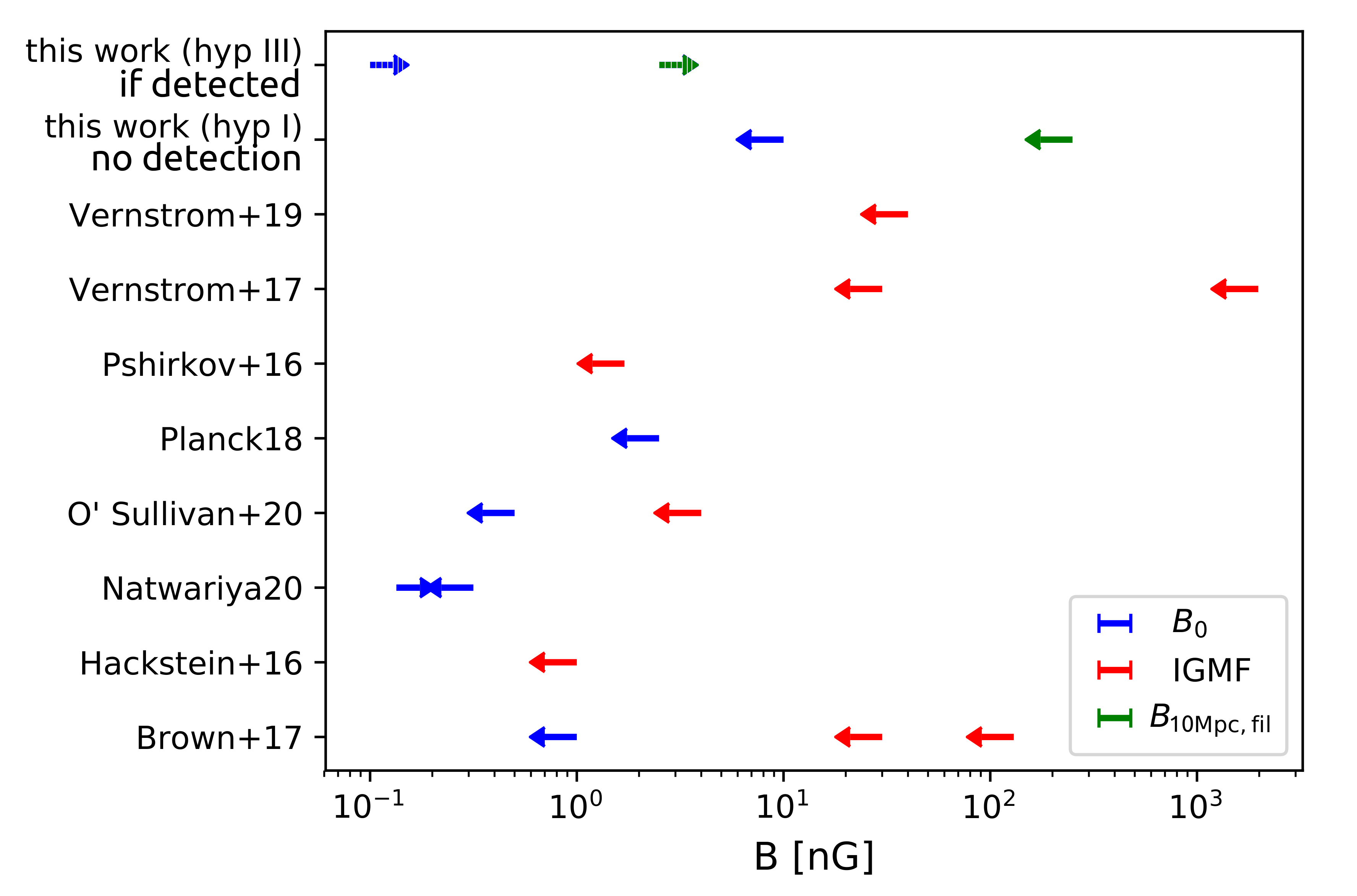}
    \caption{Summary of the current upper and lower limits to the volume-averaged IGMF (red arrows), $B$ in the filaments' WHIM (green arrows) and $B_0$ (blue arrows).}
    \label{fig:IGMF_outline}
\end{figure}

To put our new limits in comparison with other recent works (\citealt{2016MNRAS.462.3660H,2016PhRvL.116s1302P, 2017MNRAS.467.4914V, 2017MNRAS.468.4246B, 2019ApJ...878...92V, 2020MNRAS.495.2607O, 2020arXiv200709938N}, and \citealt{2019JCAP...11..028P} for joint BICEP2/Keck - Planck 2018 updated results), we show them in Fig.~\ref{fig:IGMF_outline} separating the limits inferred for the IGMF or the magnetic field intensity of the WHIM (red arrows) and for the primordial magnetic field intensity $B_0$ (blue arrows). 
We note that our limits are still in agreement with the recent limits $0.134<B_0/\rm nG<0.316$ set by the level of excess diffuse emission observed by ARCADE2 and EDGES 21cm line experiments \citep{2020arXiv200709938N}.
Furthermore, an apparent tension seems to arise between our lower limit to $B_{\rm 10Mpc}$ into filaments and the one derived from other probes such as the level of anisotropy in the arrival direction of charged ultra-high-energy cosmic rays used to limit the average amplitude of magnetic fields in voids to $\leq 1 \rm nG$ \citep{2016MNRAS.462.3660H}, or from the 
non-detection of a trend of rotation measures from distant radio sources with respect to redshifts \citep{2016PhRvL.116s1302P}. Although computed over similar linear scales $\geq \rm Mpc$ and globally refer to the IGM, they can still hardly be directly compared since referred to different IGM environments (e.g. voids, filaments, averaged).
Interestingly a recent work suggests that primordial magnetic fields with amplitude $\sim 0.1 ~\rm nG$ would possibly alleviate the existing tension between cosmological and standard candle-based estimates of $H_0$ \citep{2020arXiv200409487J}.

While it is hard to derive conclusive  limits from  these data, as no robust detection (although tentative) of the diffuse emission from the cosmic web can be claimed, this first attempt stresses the potential of low-frequency radio observations in constraining extragalactic magnetic fields, and its relevance to the study of cosmic magnetogenesis. 
With the analysis and the values obtained in this work, we can forecast to produce tighter constraint than the ones posed by CMB experiments by covering a $\sim \times 10$ larger sample of cluster pairs similar to the ones analysed here even in the case of other non-detections.

\section{Data availability}

The code used to produce the simulations discussed in this paper is public (enzo-project.org). Significant subsets of the simulations are publicly available at https://cosmosimfrazza.myfreesites.net/the\_magnetic\_cosmic\_web, while larger data set can be shared upon request

\section*{Acknowledgements}\label{acknowledgments}
\vspace{-0.2cm}
We thank Klaus Dolag for providing corrections and helpful scientific feedback.
NL, and FV acknowledge financial support from the ERC Starting Grant "MAGCOW", no. 714196. 
ABon acknowledges financial support from the ERC  Starting Grant "DRANOEL", no. 714245.
ABot acknowledges support from the VIDI research programme with project number 639.042.729, which is financed by the Netherlands Organisation for Scientific Research (NWO). 
Radio imaging made use of WSClean v2.6 \citep{2014MNRAS.444..606O} and DDF \citep{2018A&A...611A..87T}.
NL wishes to thank S. Carozzi and A. Crotti for psychological aid during and after the lock-down that followed the covid-19 pandemic situation, when this work took shape.
This paper is based (in part) on data obtained with the International LOFAR Telescope (obs. ID LC9\_020, PI F.V.) and analysed using LOFAR-IT infrastructure. 
LOFAR (van Haarlem et al. 2013) is the Low Frequency Array designed, constructed by ASTRON and collectively operated by the ILT foundation. 
These data were (partly) processed by the LOFAR Two-Metre Sky Survey (LoTSS) team. This team made use of the LOFAR direction independent calibration pipeline (https://github.com/lofar-astron/prefactor) which was deployed by the LOFAR e-infragroup on the Dutch National Grid infrastructure with support of the SURF Co-operative through grants e-infra 160022 and e-infra 160152 (\citealt{2017isgc.confE...2M}, PoS(ISGC2017)002).
The cosmological simulations were performed with the {\enzo} code (http://enzo-project.org), which is the product of a collaborative effort of scientists at many universities and national laboratories.FV acknowledges the  usage of computational resources on the Piz Daint supercomputer at CSCS-ETHZ (Lugano, Switzerland) under project s805, and he also acknowledges the usage of online storage tools kindly provided by the Inaf Astronomica Archive (IA2) initiave (http://www.ia2.inaf.it). 

\bibliographystyle{aa}
\bibliography{aa} 

\label{lastpage}
\end{document}